# Chaotic Behaviour of Atomic Energy Levels

**A. YILMAZ[a], G. HACIBEKIROGLU[a], E.I BOLCAL[a] and Y. POLATOGLU[b]**

[a]Department of Physics, TC İstanbul Kultur University, 34156 İstanbul, Turkey

[b]Department of Mathematics and Computer Science, TC İstanbul Kultur University, 34156 İstanbul, Turkey

**Abstract:**

The authors of this paper studied Schrodinger wave equation to investiagate the chaotic behavior of atomic energy levels in relation with three quantum numbers n, l, m by means of derived inequality. It could give rise to the siplitting of atomic spectral lines.

**Introduction:**

In an atomic spectra that measure radiation absorbed or emitted by electrons "jumping" from one "quantum state" to another, a quantum state is represented by values of *n, l,* and *m*. So called "selection rules" limit what "jumps" are possible. Generally a jump or "transition" is only allowed if all these three numbers change in the process. This is because a transition will be able to cause the emission or absorption of electromagnetic radiation if it involves a change in electromagnetic dipole of the atom. On the other hand, close examination reveals that some of the spectral lines display splitting as the follows,

- So-called fine structure splittting occurs because of an interaction between the spin and and motion of the outermost electron (*spin-orbit coupling*).
- Some atoms can have multiple electron configurations with the same energy level, thus appears a s single line. The interaction of the magnetic field with the atom shifts these electron configurations to slightly different energies, resulting in multiple spectral lines (*Zeeman effect*).
- The presence of an external electric field can cause a comparable splitting and shifting of spectral lines by the modifying the electron energy levels (*Stark effect*).

It was the effort to explain this radiation that led to the first successful quantum theory of atomic structure, developed by Niels Bohr in 1913. Bohr's one-dimensional model used one quantum number to describe the distribution of electrons in the atom. The only information that was important was the size of the orbit, which was described by the n quantum number. Although the Bohr theory does a good job of predicting energy levels for the hydrogenic (*one-electron*) atom , and fails even for helium, it predicts nothing about transition rates between levels.

In 1916 the fine-structure constant was introduced into physics by Arnold Sommerfeld as a measure of the relativistic deviations in atomic spectral lines from the predictions of the Bohr model. It appears naturally in Sommerfeld's analysis and determines the size of the splitting or fine-structure of the hydrogenic spectral lines. However the proof of Sommerfeld is not effective in a alkali metals.

Fine structure splitting indicates that up to two electrons can occupies a single orbital. Hovewer, two electrons can never have the same exact quantum numbers according to Hund's rule, which addresses the Pauli's exclusion principle.

In 1925 the discoveries of Goudsmit and Uhlenbeck suggested that the electron itself might have an intrinsic angular momentum that was (somehow) half as large as the smallest allowable nonzero orbital angular momentum - what we now call " *spin[1/2]* ".

In 1926 Schrödinger extended the de Broglie concept of matter waves to the wavefunction concept, by providing a formal method of treating the dynamics of physical particles in terms of associated waves. to describe quatum state of a single electron bound to the atomic nucleus by means of quantum numbers to the three-dimensional wavefunction model of the atom.

The principal quantum number (*n*), arose in the solution of the radial part of the wave equation, the azimuthal quantum number (*l*), arose in the solution of the polar part of the wave equation, and the magnetic quantum number(*m*) arose in the solution of the azimuthal part of the wave equation. Physical meaning of the quantum numbers can be explained as the followings:

The principal quantum number (*n*) (*n = 1, 2, 3, 4 ...*) describes the size of the orbital. The sets of orbitals with the same n-value are often referred to as electron shells or energy levels.

The angular quantum number (*l*) (*l = 0, 1 ... n−1*) describes the shape of the orbital and for an atomic orbital The various orbitals relating to different values of *l* are sometimes called sub-shells.

The magnetic quantum number (*m*),. ($m_l = -l, -l+1 ... 0 ... l-1, l$) represents the number of possible values for available energy levels of a subshell *l*. It determines the energy shift of an atomic orbital due to an external magnetic field, hence the name magnetic quantum number (Zeeman effect).

The spin quantum number (*s*) ($m_s = -1/2$ or $+1/2$), the intrinsic angular momentum of the electron to explain the existence of two electrons in the same orbital. It determines spin-orbit coupling effect to result in fine structure splitting. [1] [2]

All these four quantum numbers quantum numbers *n, l, m,* and *s* specify the complete and unique quantum state of a single electron in an atom called its wavefunction or orbital.

In the next section, alternatively, we proposed that the chaotic behavior of the Schördinger wavefunction related to the electron in an atom would deform the structure of energy levels from single line to multiple lines. In turn this would explain the splitting or shifting of atomic spectral lines in relation with three quantum numbers *n, l, m* through the derived inequality.

**Calculations:**

To get the nessary condition for the chaotic behavior of energy levels of an hydrogen atoms we utilize the radial Schrödinger equation of a particle of a reduced mass $\mu = \dfrac{mM}{m+M}$ moving at the potential $V(r) = -Ze^2/r$ in three dimensional space is

$$\frac{d}{dr}\left(r^2 \frac{dR}{dr}\right) + \frac{2\mu^2 r^2}{h^2}\left[E + \frac{Ze^2}{r} - \frac{l(l+1)h^2}{2\mu^2 r^2}\right] = 0 \tag{1.1}$$

The potential $V(r)$ is attractive ($V \leq 0$), then only the existence of linked states are possible when ($E < 0$). For this reason, in this paper we will study (investigate) the connected states. If $\rho$ is a variable and $\lambda$ is energy parameter, then

$$\rho = ar$$

$$\lambda = \frac{Ze^2}{h}\sqrt{-\frac{\mu}{2E}} \quad , \quad a^2 = -8\mu E/h^2 \quad , \quad E < 0 \tag{1.2}$$

by using the transformations is defined by the equation (1.2), the differential equation (1.1) is dimensionless then it can be defined as

$$\frac{d}{d\rho}\left(\rho^2 \frac{dR}{d\rho}\right) + \left(\lambda\rho + \frac{1}{4}\rho^2 - l(l-1)\right)R = 0 \tag{1.3}$$

Otherwise (by the way), the asymptotic behaviaour of the radial function is the following form

$$for\ small\ \rho\ values, \quad \lim_{r \to 0} R(r) \cong r^l \tag{1.4}$$

$$for\ big\ \rho\ values,\ (\rho \to \infty)\ \frac{d}{d\rho}(\rho^2 \frac{dR}{d\rho}) - \frac{1}{4}\rho^2 R \cong 0$$

$$r \to 0$$

the general solution of differential equation (1.4) is

$$R(\rho) = A.e^{-\rho/2} + B.e^{\rho/2} \tag{1.5}$$

When the solution function is divergent at infinity, then $B$ must be equal to zero. If we consider this asymptotic behaviours, for the wave equation $R(\rho)$ we take the following solutions into account as

$$R(\rho) = \rho^l e^{-\rho/2} L(p) \tag{1.6}$$

by using the equation (1.6) in the radial differential equation (1.3), we obtain the following differential equation

$$\rho\frac{d^2R}{d\rho^2} + (2l+2-\rho)\frac{dL}{d\rho} + (\lambda - l - 1)L = 0 \tag{1.7}$$

If we use series method to solve the differential equation (1.7), for the solution we can define a following divergent series

$$L(\rho) = a_0 + a_1\rho + a_2\rho^2 + a_3\rho^3 + ... = \sum_{k=0}^{\infty} c_k \rho^k \tag{1.8}$$

and

$$\frac{dL}{d\rho} = a_1 + 2a_2\rho + ... = \sum_k (k+1)a_{k+1}\rho^k \quad , \quad \rho\frac{dL}{d\rho} = a_1\rho + 2a_2\rho^2 + ... = \sum_k k\, a_k \rho^k$$

$$\tag{1.9}$$

$$\frac{d^2L}{d\rho^2} = 1.2\, a_2 + 2.3 a_3 \rho + ... = \sum_k (k+1)(k+2)a_{k+1}\rho^k \quad , \quad \rho\frac{d^2L}{d\rho^2} = 1.2 a_2 \rho + 2.3 a_3 \rho^2 + ...$$

$$= \sum_k k(k+1)a_{k+1}\rho^k$$

To write the equation (1.9) in the differential equation (1.3), we can find

$$\sum [k(k+1)a_{k+1}(2l+2)(k+1)a_{k+1} - ka_k + (\lambda - l - 1)a_k]\rho^k = 0 \qquad (1.10)$$

From the equation (1.10), the recurrancy relation is

$$a_{k+1} = \frac{k+l+1-\lambda}{(k+1)(k+2l+2)}a_k \qquad (1.11)$$

and given by the equation (1.8) that we take D'alembert criteria for the convergency of the solution where the coefficients provides the recurrancy relation

$$\lim_{k \to \infty} \frac{C_{k+1}}{C_k} \to \frac{k}{k^2} = \frac{1}{k} \qquad (1.12)$$

then from the equation (1.12)

$$e^\rho = 1 + \rho + \frac{\rho^2}{2!} + ... = \sum_k \frac{\rho^k}{k!}$$

by this series expansion we can obtain

$$\lim_{k \to \infty} \frac{a_{k+1}}{a_k} = \frac{k!}{(k+1)!} = \frac{1}{k+1} \cong \frac{1}{k}$$

This shows us, our solution function behaves like ($e^\rho$) series expansion. So for the infinity terms, the series solution ($\rho \to \infty$) situated for $L(\rho)$ is divergent like ($e^\rho$). It is connected to this, the wave equation $R(\rho)$

$$R(\rho) = \rho^l\, e^{-\rho/2}\, L(\rho) = \rho^l\, e^{-\rho/2} \cong \rho^l\, e^{\rho/2}$$

will be divergent. We remove the conditions of to be divergent, for the solution function has a finite terms which means it is polinomial. Consequently in the recursion relation (1.11), the numerator has been zero after by the definite ($k_{max}$) index. This means for a given $l$, for some $k = k_{max}$, the fact

$$n = (k_{max}) + l + 1 \text{ and } k_{max} = 0, 1, 2, ... \qquad (1.13)$$

takes every value on the equation (1.13), the orbital quantum number $l = n - 1 - k_{max}$ just taken $l = n-1, n-2, ..., 2, 1, 0$ values or it must be $l \leq n-1$. When we use the equation bounding $n$ parameter to the energy

$$n = \frac{Ze^2}{h}\sqrt{-\frac{\mu}{2E}} \qquad (1.14)$$

the atomic energy levels are obtained, these are

$$E_n = -\frac{1}{2}\mu c^2 \frac{(Z\alpha)^2}{n^2} \qquad (n=1,2,3,\ldots) \qquad (1.15)$$

If we set $\lambda = n$, then the recursion relation (1.11) after some calculations is

$$a_{k+1} = (-1)^{k+1} \frac{n-(k+l+1)}{(k+1)(k+2l+2)} \cdot \frac{n-(k+l)}{k(k+2l+1)} \cdots \frac{n-(l+1)}{1\cdot(2l+2)} a_0 \qquad (1.16)$$

and using the power series expansion for (1.8), we observe that this equation is called as associated Laguerre polynomials [3]

$$H(\rho) = L_{n-l-1}^{2l+1}(\rho) \qquad (1.17)$$

Therefore, the solution of (1.17) is given as

$$R_{nl}(\rho) = N_{nl}\, \rho^l\, e^{-\rho/2}\, L_{n+l}^{2l+1}(\rho) \qquad (1.18)$$

This radial wavefunction solution what we look for radial providing a relation between the orbital and angular quantum numbers.

Up to now, the answer of the following question was not given, " What kind of relation between the quantum numbers will define a chaotic behaviour of an energy levels? ".

Referring to the paper by G.Hacibekiroğlu, M.Çağlar and Y. Polatoğlu [4] and using the recursion relation (1.16), the necessary condition to behave chaotically can be expressed as

$$a_2^2 - \frac{2(n-l-1)+4}{2(n-l-1)+3} a_1 a_3 > 0 \quad \text{where } a_1 \neq 0, \text{ and}$$

$$a_1 = c_k = c_{n-l-1}$$
$$a_2 = c_{k-1} = c_{n-l-2} = (-1)^{-1}(n-l-1)(n+l)c_{n-l-1}$$
$$a_3 = c_{k-2} = c_{n-l-3} = \frac{(-2)^{-2}}{2}[(n-l-1)(n+1)][(n-l-2)(n+l-1)]c_{n-l-1}$$

After some arrangments, we found that for a given angular momentum quantum number $l$, which is permitted to take the values only up to $n-3$, a chaotic behavior achieves whenever the principle quantum number provides the condition written below:

$$n < \left[l - \frac{2}{3}(2\ell+1)(y_i - 1)\right] \qquad (1.19)$$

where

$$y_i = \begin{cases} \cosh\left(\frac{1}{3}\cosh^{-1}\kappa\right) & \text{for } \ell = 0 \\ -\cosh\left(\frac{1}{3}\cosh^{-1}\kappa\right) & \text{for } \ell = 1,2,3,\cdots,n-3 \end{cases}$$

[5] and

$$\kappa = -\left(\frac{27}{2}\right)\left[1 + 27\cdot\left(\frac{\ell-1}{2\ell+1}\right)\right]$$

Accordingly, the chaotic behavior of energy levels $E_n$ can be detemined by means of the following inequality:

$$|E_n| > (Z\alpha)^2 \left(\frac{1}{2}\mu c^2\right)^2 \left[l - \frac{2}{3}(2\ell+1)(y_i - 1)\right]^{-2}$$

As seen in (1.15) the contribution from Schrödinger wave equation to *fine structure spliting* of hydrogen atom to order $\alpha^2$. Since the nonrelativistic kinetic enery in Hamiltonian is used there. However, the expansion of relativistic kinetic energy will contribute to order $\alpha^4$. Also, the electron not only has orbital angular momentum L, intrinsic angular momentum S, so-called spin. Therefore, the energy correction for the Schrödinger's equation by combing the spin-orbit coupling effect kinetic energy correction we need.

**Conclusion:**

If one wishes to quote the energy splittings of the hydrogen atom accurate to order $\alpha^4$, the complete energy correction for fine structure to order $(Z\alpha)^4$ is also to be considered:

$$\Delta E_{fs} = \Delta E_{rel} + \Delta E_{so} = -(Z\alpha)^4 \mu c^2 \frac{1}{2n}\left[\frac{1}{\left(j+\frac{1}{2}\right)} - \frac{3}{4n}\right]$$

One can notice that the energy correction, which is called *the fine sturcture of the hydrogen atom,* depends only on the spin quantum number *j*, due to being of order $\alpha^2 \sim 10^{-4}$ times smaller than the principle energy splittings. This why $\alpha$ is known as the *fine sturcture constant*.

Previously, the theory of Dirac theory modified the Schrödinger Equation for relativistic limit. On the other hand our study an attempt to investigate the Schrödinger equation at the chaotic limit. To conclude we derive the necessary condition for the chaotic behavior as an alternative explaination some kind of for the splitting or shifting of atomic spectral lines.